\documentclass[sigconf]{acmart}
\AtBeginDocument{%
  }

\copyrightyear{2026}
\acmYear{2026}
\setcopyright{cc}
\setcctype{by-nc-nd}
\acmConference[KDD 2026] {Proceedings of the 32nd ACM SIGKDD Conference on Knowledge Discovery and Data Mining V.2}{August 9--13, 2026}{Jeju Island, Republic of Korea.}
\acmBooktitle{Proceedings of the 32nd ACM SIGKDD Conference on Knowledge Discovery and Data Mining V.2 (KDD 2026), August 9--13, 2026, Jeju Island, Republic of Korea}
\acmISBN{979-8-4007-2259-2/2026/08}
\acmDOI{10.1145/3770855.3818836}
\usepackage{multirow}
\usepackage{tcolorbox}
\tcbuselibrary{listings,breakable}

\newtcblisting{promptbox}{
  listing only,
  breakable,
  colback=gray!5,
  colframe=gray!40,
  listing options={
    basicstyle=\ttfamily\footnotesize,
    breaklines=true,
    breakatwhitespace=true
  }
}

\begin{document}
\title{Biological Reasoning-Informed Regression for \\ Interpretable Regulatory DNA Activity Prediction}

\author{Yi Duan}
\authornote{Yi Duan and Zhao Yang contributed equally to this work. This work was done while they were visiting Zhongguancun Academy.}
\email{2023200660@ruc.edu.cn}
\affiliation{%
  \institution{Gaoling School of Artificial Intelligence}
  \institution{Renmin University of China}
  \city{Beijing}
  \country{China}
}

\author{Zhao Yang}
\authornotemark[1]
\email{yangyz1230@ruc.edu.cn}
\affiliation{%
  \institution{Gaoling School of Artificial Intelligence}
  \institution{Renmin University of China}
  \city{Beijing}
  \country{China}
}

\author{Jiwei Zhu}
\email{2021201632@ruc.edu.cn}
\affiliation{%
  \institution{Gaoling School of Artificial Intelligence}
  \institution{Renmin University of China}
  \city{Beijing}
  \country{China}
}

\author{Ying Ba}
\email{yingba88@outlook.com}
\affiliation{%
  \institution{Gaoling School of Artificial Intelligence}
  \institution{Renmin University of China}
  \city{Beijing}
  \country{China}
}

\author{Chuan Cao}
\email{chuancao@bza.edu.cn}
\affiliation{%
  \institution{Zhongguancun Academy}
  \city{Beijing}
  \country{China}
}

\author{Bing Su}
\authornote{Corresponding author.}
\email{bingsu@ruc.edu.cn}
\affiliation{%
  \institution{Gaoling School of Artificial Intelligence}
  \institution{Renmin University of China}
  \city{Beijing}
  \country{China}
}

\renewcommand{\shortauthors}{Yi Duan, Zhao Yang et al.}

\begin{abstract}
DNA cis-regulatory elements (CREs) such as enhancers control gene expression levels. Accurately predicting regulatory activity from DNA sequences is valuable but challenging, as it requires understanding complex biological regulatory processes. Existing methods typically regress activity scores from sequences in a black-box manner, limiting both interpretability and regression performance. Meanwhile, large language models (LLMs) benefit from explicit reasoning processes, yet directly applying LLMs to raw DNA sequences performs poorly. In this paper, we bridge this gap by introducing R3LM, a framework that teaches LLMs reasoning-informed regression on regulatory DNA through structured biological knowledge. Specifically, we design a biologically grounded data format that structures DNA's regulatory information for improved LLM understanding, and construct CRE-ReasonBench, the first dataset that associates DNA sequences and activity scores with mechanistic reasoning traces. Through two-stage training that first teaches LLMs reasoning over structured biological information then performs regression, R3LM achieves state-of-the-art performance on enhancer prediction across three cell types, outperforming both LLMs with raw sequence input and specialized DNA models while providing interpretable mechanistic explanations. We expect R3LM as an interpretable reward model that can effectively assist biologists in CRE design. Code is available at  \url{https://github.com/DuanYi516/R3LM}.

\end{abstract}

\ccsdesc[500]{Applied computing~Computational genomics}

\keywords{genomics, dna, regulatory element, large language model, chain-of-thought, motif}

\maketitle

\section{Introduction}

DNA sequences, composed of four nucleotides (\textit{A}, \textit{C}, \textit{G}, \textit{T}), serve as the blueprint of life~\citep{enformer, nucleotide_transformer}. Among these sequences, cis-regulatory elements (CREs), such as enhancers, are short but functionally critical regions that directly control gene expression levels and thereby regulate biological processes~\citep{de2020deciphering, gosai2024machine, taskiran2024cell}. Over the past decade, millions of putative CREs have been identified~\citep{gao2020enhanceratlas}. However, these naturally occurring CREs may not be optimal for human needs. The ability to design CREs tailored to specific requirements would be valuable for metabolic engineering~\citep{vaishnav2022evolution, taco}, personalized medicine~\citep{goetz2018personalized}, and smart agriculture~\citep{lorenz2011genomic}.

Traditional CRE design relies heavily on expert biological knowledge. CREs typically function through short sequence motifs, known as transcription factor binding sites (TFBSs), which recruit specific transcription factors (TFs) to activate or repress target genes~\citep{de2024targeted}. Experts design CREs by manually arranging known TFBSs based on their regulatory effects~\citep{de2025modelling}. While this approach is grounded in mechanistic understanding, it explores only a limited design space and struggles to discover CREs at scale. An alternative approach, directed evolution~\citep{wang2021directed}, generates libraries of sequence variants through mutagenesis or recombination and iteratively screens or selects variants with desired regulatory activity in wet-lab assays. Although this strategy can identify functional CREs, it remains labor-intensive, time-consuming, and costly, making broad and iterative exploration of the CRE design space difficult.

Computational methods offer a promising alternative. If we can accurately predict CRE activity directly from sequence, we can use these predictions to guide sequence optimization in silico, enabling rapid exploration of vast design spaces without experimental synthesis. In this paradigm, the prediction model serves as a \textit{reward model}~\citep{ba2025,rrm}—
providing activity scores that guide optimization algorithms such as genetic algorithms~\citep{de2020deciphering}, gradient-based search~\citep{gosai2024machine}, or reinforcement  learning~\citep{taco, ctrldna}.

Recent advances in high-throughput assays, particularly massively parallel reporter assays (MPRAs)~\citep{mrpa}, have enabled the generation of large-scale datasets pairing CRE sequences with their measured activities~\citep{de2020deciphering, vaishnav2022evolution, gosai2024machine}. Several studies~\citep{reglm, reddy2024designing, vaishnav2022evolution, gosai2024machine} have leveraged these datasets to train end-to-end neural networks that predict regulatory activity from sequence. These models have successfully guided functional CRE design, with experimental validation in wet-lab~\citep{de2024targeted, reddy2024designing}. Despite operating as black boxes, post-hoc analyses~\citep{de2024targeted, reglm} show they attend to biologically meaningful features such as TFBSs, indicating they capture aspects of CRE regulatory grammar. 

However, these purely data-driven models, which lack grounding in biological mechanisms, face two critical limitations. First, when used as reward models for sequence optimization, they are vulnerable to reward hacking~\citep{taco, ctrldna, liomni}. Second, without explicit reasoning processes, these black-box predictions lack the evidence necessary for biologists to validate and trust. This raises a fundamental question: \textit{can computational models reason about CRE activity in the same way biologists do, for example by explicitly analyzing transcription factor motifs, their arrangements, and the underlying regulatory logic?}

Such reasoning capabilities are naturally associated with large language models (LLMs), which have demonstrated remarkable performance on tasks requiring step-by-step reasoning, such as mathematics and coding, where chain-of-thought (CoT)~\citep{wei2022cot} processes have been shown to improve final outcomes. However, DNA, as the \textit{language of life} is not directly comprehensible to LLMs trained on natural language. As we demonstrate in Section~\ref{sec:prelim_raw_dna}, directly asking an LLM to predict CRE activity from raw nucleotide strings yields performance far inferior to specialized black box models~\citep{enformer}, suggesting a fundamental representation gap between DNA sequences and what LLMs require.

To address this gap, we introduce the Regulatory Context Card (RCC), a structured input format that compiles raw DNA sequences into interpretable, biologically grounded representations. RCC extracts key regulatory features, including TFBSs, sequence statistics, and grammar tags derived from motif configurations—using established biological knowledge and bioinformatics tools. This transformation provides LLMs with the structured information necessary to reason about regulatory mechanisms. Furthermore, to teach LLMs to map these biological features to regulatory activity through explicit reasoning, we construct CRE-ReasonBench, the first dataset pairing DNA sequences with step-by-step mechanistic rationales that explain how motif configurations, spacing, and cellular context collectively determine activity levels. Using this dataset, we develop a two-stage training framework that first teaches an LLM to generate mechanistic reasoning traces, and then trains a regression model to predict continuous activity scores conditioned on these reasoning processes.

Through extensive experiments across enhancers of three cell types, we demonstrate that LLMs, when equipped with structured biological representations and reasoning supervision, achieve state-of-the-art CRE prediction performance, surpassing specialized black-box models~\citep{enformer}. Critically, our approach provides biologically grounded, interpretable explanations alongside predictions, enabling the model to serve as an interpretable reward model to assist biologists in designing CREs. We name our framework R3LM (\textbf{R}easoning \textbf{R}egression \textbf{R}eward \textbf{L}anguage \textbf{M}odel).

Our contributions are summarized as follows:

\begin{itemize}

\item 
We introduce RCC, a structured format that transforms raw DNA into interpretable biological representations accessible to LLMs.
\item 
We construct CRE-ReasonBench, the first dataset containing mechanistic reasoning traces for regulatory DNA activity prediction.
\item 
We propose a two-stage training framework that enables LLMs to perform reasoning-based regression on CRE activity.
\item 
Through extensive experiments, we demonstrate that biological reasoning-informed LLMs achieve SOTA performance while providing interpretable mechanistic explanations.
\end{itemize}
\section{Related Work}

\textbf{CRE activity prediction.} 
Early work by \citet{de2020deciphering} modeled CRE activity using biologically motivated, additive formulations that explicitly decomposed regulatory function into transcription factor binding and chromatin accessibility components. These mechanistic models demonstrated that sequence-to-activity relationships are predictable from first principles. However, owing to their linear structure, such models are inherently limited in their capacity to capture higher-order interactions~\citep{georgakopoulos2023transcription, dudnyk2024sequence} and complex regulatory grammar. With the rapid development of deep learning, subsequent studies began to directly learn end-to-end mappings from DNA sequence to CRE activity. Notably, \citet{vaishnav2022evolution}, \citet{de2024targeted}, and \citet{gosai2024machine} trained neural network models that predict quantitative activity scores directly from sequence. These approaches substantially improve predictive performance by capturing nonlinear effects without relying on explicitly specified regulatory assumptions. In parallel, regLM~\citep{reglm} explored transfer learning for CRE activity prediction by fine-tuning Enformer~\citep{enformer}, a representative sequence-to-function model~\citep{yang2025space, avsec2026advancing} pretrained on thousands of functional genomics tracks. Benefiting from Enformer’s large-scale pretraining, this approach demonstrates improved performance when adapted to CRE activity prediction via task-specific fine-tuning. Similarly, \citet{reddy2024designing} adopted an Enformer-initialized model and fine-tuned it on task-specific data. In our experiments, Enformer also serves as the strongest specialized baseline. However, when such black-box predictors are used in practice to guide CRE design \citep{taco, ctrldna, reddy2024designing}, they may suffer from reward hacking, while also lacking interpretability.

\textbf{LLM Reasoning. }
Chain-of-thought prompting shows that emitting intermediate reasoning steps can elicit multi-step reasoning in LLMs and provides an inspectable trace \cite{wei2022cot,kojima2022zeroshot,wang2022selfconsistency}.
Complementarily, RL-based training can explicitly incentivize reasoning behaviors in LLMs \cite{deepseek2025}.

\textbf{LLM for DNA understanding}
A growing line of work interfaces biological sequence models with LLMs to enable natural-language querying and mechanistic reasoning.
BioReason tightly integrates a DNA foundation model with an LLM and trains it to produce multimodal biological deductions \cite{bioreason}. ChatNT projects a DNA encoder's representations into an English decoder's token embedding space to solve biological sequence tasks in an instruction-following form \cite{chatnt}. In reward modeling, however, rationales may become post-hoc narratives that are weakly coupled to the predicted score. 
We advance this by introducing a compiled schema and anchor-token readout, which condition continuous regression on explicit reasoning traces to ensure the rationale remains causally informative.
\begin{figure*}[t]
  \centering
  \includegraphics[width=\textwidth]{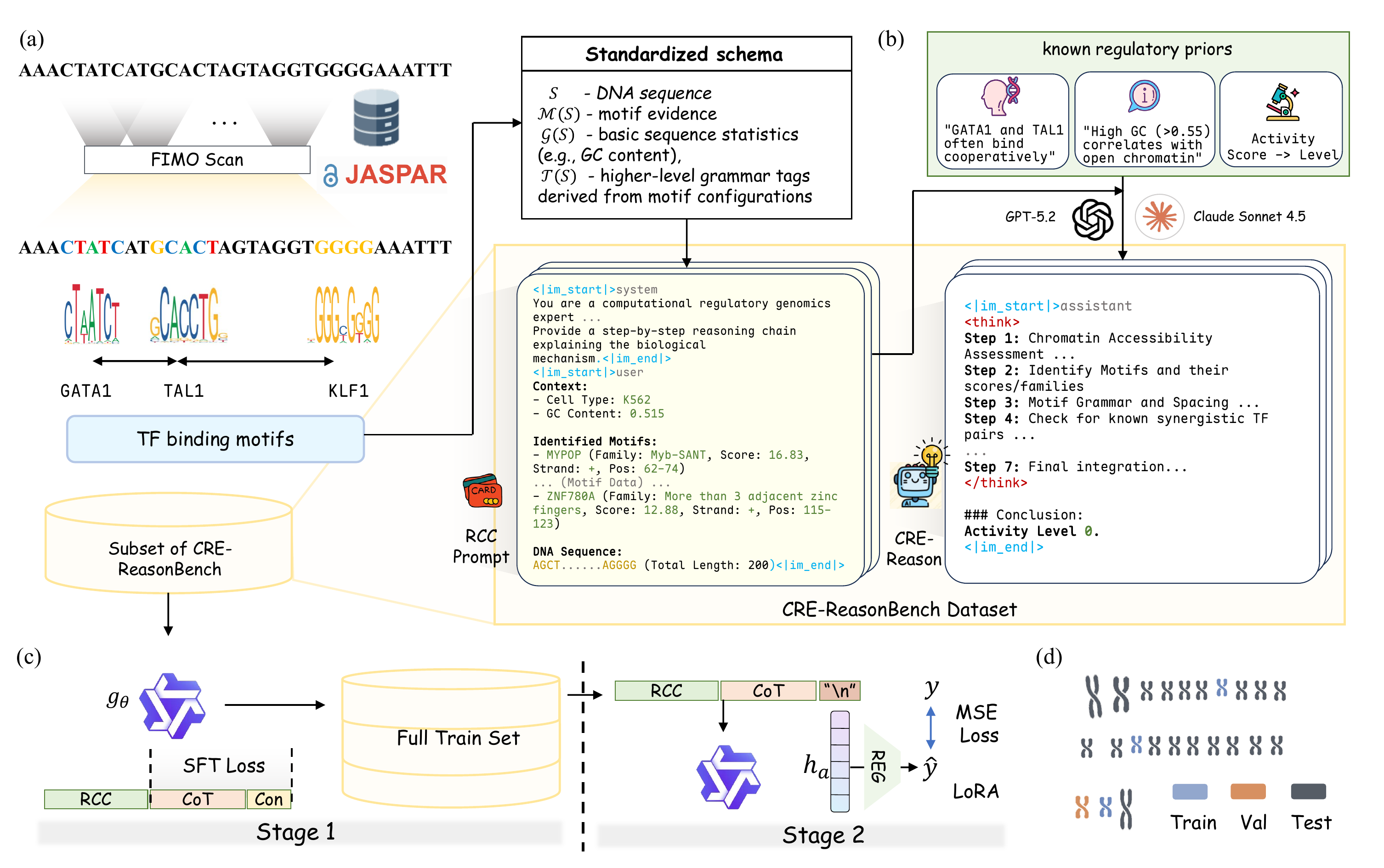}
  \caption{Overview of R3LM framework. (a) Assembly of the RCC in Section~\ref{sec:rcc}.
(b) Construction of the CRE-ReasonBench dataset in Section~\ref{sec:reasonbench}.
(c) Two-stage training procedure of the generative reward model in Section~\ref{sec:training}.
(d) Chromosome-based train/validation/test split used in experiments.}
  \Description{Overview of R3LM framework.}
  \label{fig:overview}
\end{figure*}

\section{Method}
We first show that raw DNA strings are a poor interface for LLM reasoning in Section \ref{sec:prelim_raw_dna}. We then introduce RCC in Section~\ref{sec:rcc}. Next, we introduce CRE-ReasonBench in Section~\ref{sec:reasonbench}. Finally, in Section \ref{sec:training} we propose a two-stage training strategy.

\subsection{Preliminary Study}
\label{sec:prelim_raw_dna}

A central challenge in applying instruction-tuned LLMs to CRE modeling is that the canonical input is a long nucleotide string (e.g., ``\texttt{ACGT...}''), which provides little structure that an LLM can readily operate on. This issue has been noted in recent discussions of genomic reasoning: when DNA is presented as plain text, LLMs tend to treat it as an arbitrary string and fail to recover biologically meaningful features without additional scaffolding (e.g., motif identity, spacing, and context) \cite{bioreason}. Motivated by this, we conduct a targeted probe to quantify how much a strong instruction-tuned LLM can infer \emph{directly} from raw sequences, and whether injecting lightweight, biologically grounded context can substantially improve this.

\textbf{Probe design.}
Directly asking an LLM to predict an absolute activity value or binned level from a nucleotide string is a substantially harder problem. 
In contrast, we evaluate a modern instruction-tuned LLM in two controlled, pairwise tasks where random guessing yields $50\%$ accuracy. Both tasks are intentionally simple in format (two candidates per query) to isolate whether the model can map nucleotide strings to regulatory function, rather than testing long-horizon planning or tool use.
We implement this probe by querying GPT-5.2 via the OpenAI API and computing accuracy from the forced-choice outputs.

\textbf{Task 1: Cell-type discrimination.}
~Given two sequences, where for a given cell type only one is highly active (one HepG2-specific and one K562-specific), the model is asked to determine which sequence corresponds to which cell type (or equivalently, which is more active in HepG2 vs.\ K562). This task probes whether an LLM can infer cell-type-specific regulatory cues from raw sequence alone.

\textbf{Task 2: Within-cell ranking (relative activity).}
Given two sequences from the \emph{same} cell type, where one is empirically high-activity and the other low-activity, the model is asked to predict which sequence has higher regulatory activity. This task probes whether the model can perform a minimal form of functional comparison from raw sequence strings.

\textbf{Results.}
Table~\ref{tab:motivating_raw_dna} summarizes the results. When prompted zero-shot with \emph{raw DNA only}, the LLM performs at the level of random on cell-type discrimination and below chance on within-cell ranking, suggesting that the model fails to consistently extract functional regulatory signals from unstructured nucleotide strings.

In contrast, fine-tuned domain models like Enformer \cite{enformer} outperform LLMs by encoding regulatory priors that are inaccessible via naive string prompts. However, providing LLMs with explicit features (e.g., motifs and GC content) bridges this gap, boosting ranking performance from $38\%$ to $74\%$. This suggests the bottleneck is a representation mismatch between raw DNA and regulatory logic, rather than the LLM's reasoning ability. Consequently, we propose a standardized schema (Section~\ref{sec:rcc}) that compiles DNA into structured observations to enable effective reasoning and regression.
\begin{table}[t]
\centering
\small
\begin{tabular}{lcc}
\toprule
\textbf{Model / Input} & \textbf{Cell-type discr.} & \textbf{Within-cell rank} \\
 & (\% $\uparrow$) & (\% $\uparrow$) \\
\midrule
Enformer \cite{enformer} & 68 & 94 \\
LLM (raw DNA sequence only) & 49 & 38 \\
LLM (+ motif \& GC context) & -- & 74 \\
\bottomrule
\end{tabular}
\caption{\textbf{Raw-sequence probing.} Zero-shot accuracy (\%) on two pairwise probes (chance $=50\%$). A strong instruction-tuned LLM fails to reliably compare or classify raw nucleotide strings, but improves substantially when provided with lightweight, biologically grounded context. ``--'' indicates that we focus the context-augmentation analysis on the within-cell ranking probe, which most directly reflects functional comparability.}
\label{tab:motivating_raw_dna}
\end{table}

\subsection{Regulatory Context Card (RCC)}
\label{sec:rcc}

The probe in Section~\ref{sec:prelim_raw_dna} suggests a clear bottleneck: presenting cis-regulatory DNA as a raw nucleotide string creates a \emph{representation mismatch} for instruction-tuned LLMs. While CRE function is governed by structured latent factors (e.g., transcription factor binding motifs, family-level logic, spacing/grammar, and experimental context), these factors are not explicit in ``\texttt{ACGT...}'' tokens. To bridge this gap, we introduce the RCC, a compiled and verifiable input schema that transforms each raw sequence into a standardized set of biologically grounded intermediate factors.

\textbf{Definition.}
Given a DNA sequence $S$ and experimental context $c$ (e.g., cell type), RCC is a deterministic transformation
\begin{equation}
\mathrm{RCC}(S, c) \;=\; \big[\, \mathcal{M}(S),\;\mathcal{G}(S),\;\mathcal{T}(S),\;c,\;S \,\big],
\end{equation}
where $\mathcal{M}(S)$ denotes motif evidence, $\mathcal{G}(S)$ denotes basic sequence statistics (e.g., GC content), and $\mathcal{T}(S)$ denotes higher-level grammar tags derived from motif configurations. Each component is produced by programmatic procedures, making the resulting prompt \emph{auditable} and \emph{reproducible}.

\textbf{RCC construction pipeline.}
RCC is constructed in five steps, designed to expose the key factors that mechanistically mediate cis-regulatory activity while retaining traceability to the underlying sequence.

\textbf{(1) Motif evidence extraction.}
We scan each sequence against JASPAR~\cite{2026jaspar} using a standard motif scanning procedure, retaining statistically significant matches under a fixed threshold \cite{2026jaspar,fimo}. Each retained match is normalized into a tabular record containing: TF family (or TF identifier), genomic location, strand, and match score. 

\textbf{(2) Sequence statistics profiling.}
We compute simple, comparable statistics from $S$ such as GC content, which serves as a practical proxy for CpG/CGI-linked chromatin and transcriptional properties relevant to enhancer activity prediction~\cite{CpG}.
These statistics provide global priors that modulate regulatory interpretation and help the model calibrate expectations.

\textbf{(3) Grammar heuristics and grammar tags.}
Beyond individual motifs, CRE function often depends on motif \emph{arrangements} (co-occurrence, clustering, spacing, and putative composite elements). We therefore derive lightweight grammar tags $\mathcal{T}(S)$ using rule-based heuristics operating on the motif table. These tags do not attempt to fully reconstruct causal regulatory grammar; rather, they provide structured hypotheses that are explicit, inspectable, and useful for stepwise reasoning.

\textbf{(4) Experimental context injection.}
We represent assay- and condition-specific factors as explicit RCC fields. This design makes cross-context evaluation and cell-type specificity naturally expressible to the model, rather than relying on implicit prompt conventions.

\textbf{(5) Schema assembly.}
Finally, we assemble all fields into a fixed, standardized schema with consistent headings and ordering (Figure~\ref{fig:overview}). This standardization is crucial: it enforces a shared observation model across samples, enabling controlled ablations and stable training for reasoning-conditioned regression.

We provide further analysis of why RCC helps in details in Appendix \ref{app:why_rcc}.

\subsection{CRE-ReasonBench}
\label{sec:reasonbench}

A second barrier to reasoning-based reward modeling in regulatory genomics is that most datasets provide \emph{only} sequence-level labels such as activity scores, with no mechanistic explanation. However, beyond regressing scalar activity scores, we aim to develop a predictor whose output is \emph{conditioned on an explicit reasoning trace}. This requires paired supervision of the form \texttt{(RCC prompt)} $\rightarrow$ \texttt{(mechanistic rationale + label-consistent conclusion)}.

\textbf{Dataset overview.}
We therefore construct \emph{CRE-ReasonBench}, a rationale-augmented dataset of nearly 83k samples built on top of RCC prompts and observed activity levels on a discrete scale (0--3) following regLM\cite{reglm}. Each example consists of: (i) an RCC prompt derived from $(S,c)$, (ii) a stepwise mechanistic rationale grounded in the RCC evidence, and (iii) a label-consistent conclusion tied to the observed activity.
Figure~\ref{fig:overview} provides an end-to-end visualization of the generation pipeline and summarizes key dataset statistics (composition, split protocol, and rationale characteristics) that are typically reported in a dataset table.

\textbf{Rationale synthesis protocol.}
To synthesize expert-like rationales, we condition a frontier LLM on (a) the RCC prompt, (b) the observed activity label, and (c) a constrained instruction protocol that enforces mechanistic attribution. Concretely, the instruction protocol requires the model to:
(i) produce a fixed number of causal steps (e.g., 5--7),
(ii) explicitly reference RCC evidence (motifs, families, grammar tags, and GC content) in each step,
(iii) reason from global priors (e.g., GC content) to local motif evidence, then to grammar/spacing hypotheses and synergy,
and (iv) end with a short conclusion that matches the observed activity level on the 0--3 scale.
An example RCC prompt and its synthesized rationale are shown in Figure~\ref{fig:overview}(a).

\textbf{Structured format and automatic validity checks.}
Because the rationales will later be used both as supervision for a \emph{CoT generator} and as conditioning inputs for a \emph{reason-conditioned regressor}, CRE-ReasonBench uses a strictly parseable output format (e.g., a delimited \texttt{<think>} block plus a final conclusion line). We apply automatic checks to ensure:
(a) schema compliance (required headers and delimiters),
(b) step structure validity (presence/ordering of numbered steps),
and (c) conclusion-field presence and parseability.
These checks enable robust downstream training without manual intervention and support reproducible filtering policies (e.g., filtering by format validity only).

\textbf{Intended use: reasoning supervision and oracle evaluation.}
CRE-ReasonBench serves two roles in our framework.
First, it provides high-quality supervision to train a CoT generator that learns to produce mechanistic rationales from RCC prompts (Section~\ref{sec:training}).
Second, the synthesized rationales define a \emph{reference} reasoning trace that we use in an \emph{Oracle-CoT} evaluation setting to isolate the regression module's capacity under correct reasoning (Section~\ref{sec:main_results}).
Importantly, while the rationale synthesis conditions on observed labels to ensure label-consistent mechanistic attribution, our downstream regression training and data construction policies avoid label-conditioned selection by filtering only for syntactic validity (Section~\ref{sec:training}), preventing the training distribution from being artificially restricted to ``easy'' or ``already-correct'' samples.

\textbf{Remark.}
CRE-ReasonBench does not claim to provide ground-truth causal explanations; rather, it provides \emph{grounded and auditable} mechanistic rationales that are constrained by programmatic evidence (RCC fields) and known regulatory priors. This form of structured explanation is precisely what we require to train a generative reward model whose scalar predictions are explicitly linked to intermediate, inspectable factors.

\subsection{Two-Stage Training}
\label{sec:training}

Our objective is to train a CRE activity prediction model that simultaneously (i) produces an explicit mechanistic rationale grounded in RCC evidence and (ii) outputs a \emph{continuous} reward aligned with quantitative activity measurements. This ``reasoning-conditioned regression'' perspective is closely related to recent regression-aware chain-of-thought training recipes, which emphasize that reasoning traces can be treated as intermediate factors that improve scalar prediction and that training should account for the mismatch between reference rationales and model-generated rationales at deployment \cite{tract}. In our setting, we realize these principles with a practical two-stage pipeline that avoids intrusive architecture changes and remains compatible with standard LLM fine-tuning infrastructures.

A key detail is that our dataset contains two related supervision signals. The underlying measurement is a continuous activity score $y\in\mathbb{R}$, which is the quantity we ultimately want as a reward for downstream optimization. For rationale supervision, however, it is substantially more natural to phrase conclusions in a small number of discrete bins (e.g., ``inactive'' to ``highly active'') that an LLM can express consistently in text. We therefore discretize $y$ into a discrete activity level $\ell\in\{0,1,2,3\}$ via a fixed thresholding rule $\ell=\tau(y)$. The crucial point is that $\ell$ is used only to teach the model how to \emph{verbalize} a label-consistent conclusion during rationale generation, whereas the regression component is trained to fit the true continuous score $y$.

While joint optimization of a language modeling objective and an anchor-based MSE loss is theoretically viable, we identify two critical bottlenecks. First, backpropagating regression gradients through long reasoning traces incurs prohibitive memory costs, limiting effective batch sizes. Second, the competing objectives of generation and regression induce gradient conflicts~\cite{liu2021cagrad}, leading to optimization instability where improvements in one task degrade the other. Consequently, we decouple these objectives into a robust two-stage design to ensure both computational efficiency and training stability.

\textbf{Stage 1: learning to generate mechanistic rationales.}
We train a chain-of-thought generator $g_{\theta}$ on a small subset of CRE-ReasonBench containing RCC prompts paired with synthesized mechanistic rationales. The input is an RCC prompt $x=\mathrm{RCC}(S,c)$; the target output is a formatted \texttt{<think>} block containing stepwise reasoning, followed by a short conclusion that states the discrete activity level $\ell=\tau(y)$. We optimize the standard teacher-forced language-modeling objective,
\begin{equation}
\mathcal{L}_{\mathrm{cot}}(\theta)
\,=\,
-\sum_{t=1}^{T}\log p_{\theta}\!\left(z_t\,\middle|\,x,z_{<t}\right),
\end{equation}
where $z_{1:T}$ denotes the target token sequence (rationale plus a level conclusion). Training the generator to output $\ell$ rather than the raw score $y$ improves textual consistency and aligns with typical instruction-following behavior, while keeping the reasoning trace grounded in RCC evidence.

\textbf{Stage 2: learning a reason-conditioned regressor for the continuous score.}
After stage 1, we apply $g_{\theta}$ to each RCC prompt $x$ in the training split to produce a self-generated rationale $\tilde r = g_{\theta}(x)$. We then construct the regression input by concatenating the RCC prompt and the generated rationale, but \emph{explicitly removing} the final conclusion line that contains the discrete level $\ell$. This design ensures that the regression model never receives an explicit level token during training or inference; it must predict the continuous score from RCC evidence and the generated mechanistic reasoning trace.

Let $\mathrm{concat}(x,\tilde r)$ be the concatenated RCC prompt and generated rationale (with the conclusion removed). We feed it into a backbone LLM with trainable LoRA adapters and extract the hidden state at a designated anchor position $a$. Denoting this representation by $h_a\in\mathbb{R}^d$, the predicted activity score is
\begin{equation}
\hat y \,=\, f_{\Theta}(x,\tilde r) \,=\, \mathrm{Proj}(h_a),
\end{equation}
where $\Theta$ includes the regression head and LoRA parameters. We minimize
\begin{equation}
\mathcal{L}_{\mathrm{reg}}(\Theta)
\,=\,
\mathbb{E}_{(x,y)\sim\mathcal{D}}
\left[\big(f_{\Theta}(x,\tilde r(x)) - y\big)^2\right].
\end{equation}

\textbf{Self-generated rationales and format-only filtering.}TRACT~\citep{tract} demonstrates that training should reflect inference-time conditions: if the regressor will consume \emph{model-generated} rationales at deployment, training it on only reference/gold rationales can create a harmful distribution mismatch. We therefore train the regressor on self-generated rationales $\tilde r(x)$ produced by $g_{\theta}$. To ensure robustness without introducing label-conditioned biases, we filter generated outputs \emph{only} by syntactic validity (e.g., presence and well-formedness of the \texttt{<think>} block and required delimiters), and we do \emph{not} filter by whether the generator's predicted level matches the ground-truth discretized label. In our experiments, the stage-1 generator attains $0.4184$ exact accuracy on level prediction over the large training split (Enformer: $0.5776$) and $0.7936$ accuracy within $\pm 1$ level; nevertheless, we retain imperfect rationales to reflect deployment conditions. This format-only policy preserves $97.69\%$ of samples, indicating that stage 1 reliably enforces structural constraints while avoiding the pitfalls of correctness-based selection.

Taken together, this two-stage recipe yields a predictive model that is explicitly \emph{reasoning-conditioned}: it learns to generate mechanistic rationales from RCC evidence and to map those rationales to a continuous reward. The design preserves the intended semantics (regression based on reasoning), aligns training with inference-time rationale distributions, and avoids the engineering and optimization pitfalls of joint training (activation memory blow-up and gradient conflict) without requiring non-standard model architectures or training infrastructure.

\section{Experiments}

\begin{table*}[htbp]
\centering
\caption{Performance comparison of regulatory activity regression across K562, SK-N-SH, and HepG2. We benchmark genomic baselines (e.g., Enformer, NT) against LLMs using raw sequences versus our Regulatory Context Card (RCC). For our proposed R3LM, we report results using both gold-standard (Oracle-CoT) and self-generated (Generated-CoT) rationales.}

\label{tab:main_results}
\resizebox{0.8\textwidth}{!}{%
\begin{tabular}{lcc|cc|cc}
\toprule
 & \multicolumn{6}{c}{\textbf{Regression Evaluation}} \\
\cmidrule(lr){2-7}
 & \multicolumn{2}{c}{\textbf{K562}} & \multicolumn{2}{c}{\textbf{SK-N-SH}} & \multicolumn{2}{c}{\textbf{HepG2}} \\
\cmidrule(lr){2-3} \cmidrule(lr){4-5} \cmidrule(lr){6-7}
\textbf{Model} & $\rho$$\uparrow$ & RMSE$\downarrow$ & $\rho$$\uparrow$ & RMSE$\downarrow$ & $\rho$$\uparrow$ & RMSE$\downarrow$ \\
\midrule
\multicolumn{7}{l}{\textit{Pure Sequence-based Models}} \\
\cmidrule(lr){1-7}
DNA-Bert2 \cite{dnabert2} & 0.6478 & 0.8773 & 0.6818 & 0.7598 & 0.6858 & 0.7181 \\
Enformer \cite{enformer} & 0.7990 & 0.7398 & \textbf{0.8339} & 0.6467 & 0.8001 & 0.6536 \\
NT \cite{nucleotide_transformer} & 0.7450 & 0.8001 & 0.6991 & 0.8482 & 0.7086 & 0.7594 \\
\midrule
\multicolumn{7}{l}{\textit{Instruction-tuned LLM Models}} \\
\cmidrule(lr){1-7}
Qwen (Seq only) & 0.3919 & 1.1430 & 0.4944 & 1.0736 & 0.4369 & 0.9790 \\
Qwen (RCC; no CoT) & 0.6886 & 0.8715 & 0.6921 & 0.8559 & 0.6975 & 0.7690 \\
\midrule
\multicolumn{7}{l}{\textit{R3LM (Ours)}} \\
\cmidrule(lr){1-7}
R3LM (Format-only; \textbf{Oracle-CoT}) & \textbf{0.8361} & \textbf{0.6206} & 0.8140 & \textbf{0.6408} & \textbf{0.8109} & \textbf{0.6475} \\
R3LM (Format-only; Generated-CoT) & 0.8246 & 0.6782 & 0.7927 & 0.6821 & 0.8025 & 0.6771 \\
R3LM (Oracle-gated; Oracle-CoT)  & 0.8992 & 0.5746 & 0.8844 & 0.4671 & 0.9111 & 0.5047 \\
\bottomrule
\end{tabular}%
}
\end{table*}
\label{sec:experiments}

We evaluate our R3LM along three axes: (i) \emph{continuous} activity-score regression quality as a reward model, (ii) robustness to the reasoning distribution at inference (gold vs.\ self-generated rationales), and (iii) the contribution of our compiled RCC interface and modeling design choices. Unless noted otherwise, we report results on a random subset of 100 sequences sampled from the held-out chromosome split.

We first introduce our experiment setup in Section \ref{sec:exp_setup}, followed by quantitative regression results in Section 4.2. We then analyze the stage-1 reasoning generator in Section 4.3 and present ablation studies and additional evaluations in Sections 4.4 and 4.5.

\subsection{Experimental Setup}
\label{sec:exp_setup}

\begin{figure}
  \centering
  \includegraphics[width=\linewidth]{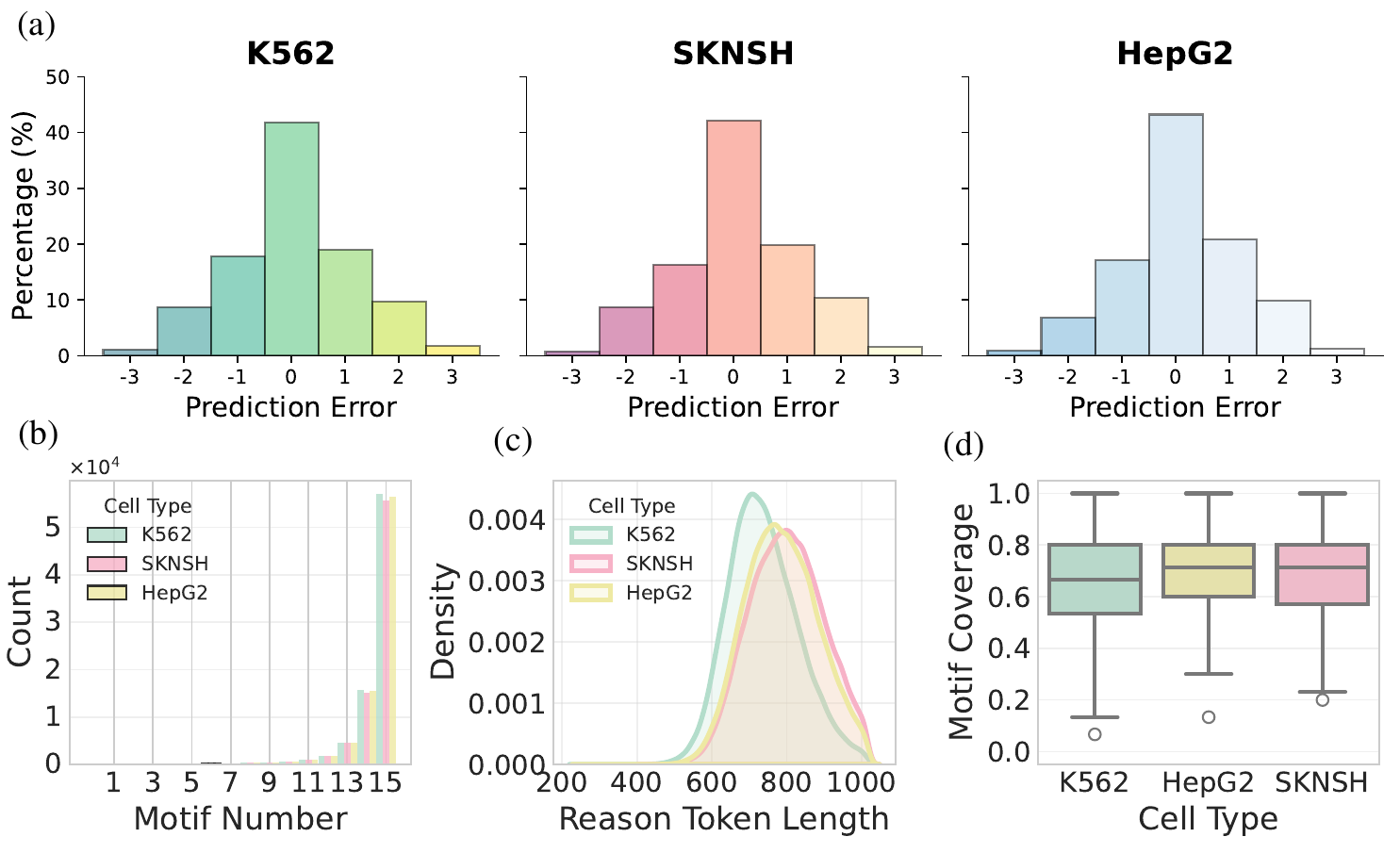}
\caption{\textbf{Stage-1 generator diagnostics and CRE-ReasonBench statistics across three cell types.}
(a) Distribution of discrete activity level prediction error $\Delta=\hat{\ell}-\ell$ for the stage-1 reasoning generator.
(b) Motif count distribution in the RCC prompts.
(c) Distribution of rationale tokens length.
(d) Motif coverage of rationales, i.e., the fraction of RCC prompt motifs explicitly referenced in the generated rationale.}
\Description{Stage-1 generator diagnostics and CRE-ReasonBench statistics across three cell types.}
\label{fig:analyze}
\end{figure}

\textbf{Datasets.}
We evaluate on human enhancer benchmarks used in prior CRE design work \cite{gosai2024machine,promoter_dataset}. The enhancer benchmark contains 200bp candidate regulatory sequences measured by massively parallel reporter assays (MPRAs) \cite{mrpa} across three cell types (HepG2, K562, and SK-N-SH). Each example provides a DNA sequence and a continuous activity \emph{score} $y\in\mathbb{R}$ for the corresponding cell type; we additionally derive a discretized activity \emph{level} $\ell\in\{0,1,2,3\}$ by thresholding $y$ (Section~\ref{sec:training}). The chromosome split protocol is summarized in Figure~\ref{fig:overview}, and dataset statistics are summarized in Figure~\ref{fig:analyze}. Following regLM \cite{reglm} regLM split, evaluation is performed on a fixed random subset of 100 sequences sampled from the held-out test chromosomes.

\textbf{Preprocessing and motif processing.}
We adopt the preprocessing pipeline described in \cite{reglm,ctrldna} to ensure comparability with prior studies. In particular, we obtain human-specific position probability matrices (PPMs) and motif metadata from JASPAR 2026 \cite{2026jaspar}. We scan each sequence for motif occurrences using FIMO \cite{fimo} and retain statistically significant hits with $p<5\times10^{-5}$. We then normalize each hit into a \texttt{MotifHit} record, sort hits by genomic position, and remove redundant overlaps: if multiple motifs from the same TF family overlap, we keep only the highest-scoring hit. This yields a compact motif evidence table with roughly 10--15 motif hits per sequence. We further annotate each hit with TF family/class and strand information via motif metadata (e.g., TRANSFAC-style annotations), which are used to assemble RCC prompts (Section~\ref{sec:rcc}).

\textbf{Models and baselines.}
We compare against (i) \textbf{DNABert2} \cite{dnabert2}, fine-tuned for regression with a lightweight head; (ii) \textbf{Enformer} \cite{enformer}, fine-tuned via transfer learning as a cell-type-specific reward model following protocols from prior CRE design studies \cite{reglm,taco,promoter_dataset,gosai2024machine}; 
(iii) a nucleotide foundation model baseline (\textbf{NT}) \cite{nucleotide_transformer} adapted to regression with a lightweight head; and (iv) an instruction-tuned LLM baseline (\textbf{Qwen}) \cite{qwen3technicalreport} fine-tuned to regress activity scores from prompts without explicit reasoning (details below and in Appendix~\ref{app:training_details}).
Our method, \textbf{R3LM}, follows the two-stage pipeline in Section~\ref{sec:training}: a CoT generator $g_\theta$ produces mechanistic rationales conditioned on RCC, and a reason-conditioned regressor $f_\phi$ predicts the continuous score from \texttt{RCC + (generated rationale)}.
We use Qwen3-Instruct~\cite{qwen3technicalreport} as the backbone for  R3LM and all Qwen-based baselines to ensure fair comparisons.

\textbf{Evaluation protocol: Oracle-CoT vs.\ Generated-CoT.}
Because R3LM is explicitly reasoning-conditioned, we evaluate it under two complementary settings.
In \textbf{Oracle-CoT}, the regressor consumes gold rationales from CRE-ReasonBench to estimate the regression module's performance under correct/idealized reasoning.
In \textbf{Generated-CoT}, we feed self-generated rationales from $g_\theta$, reflecting the deployment scenario in which the reward model must rely on its own reasoning traces.
Unless stated otherwise, all regression models are trained on self-generated rationales with \emph{format-only} filtering (Section~\ref{sec:training}).

\textbf{Evaluation Metrics.}
We primarily report Pearson correlation ($rho$) and RMSE between predicted scores $\hat y$ and ground-truth scores $y$ on the test split. This choice is standard in quantitative sequence-to-function prediction and MPRA modeling, and is also well aligned with reward modeling: in downstream optimization, the reward is only identifiable up to an affine transformation (shift and positive scaling), and correlation is invariant to such transformations.
\subsection{Performance of Activity Prediction}
\label{sec:main_results}

Table~\ref{tab:main_results} reports Pearson correlation and RMSE on the three enhancer cell types. R3LM with \emph{format-only} filtering is competitive with strong discriminative baselines and achieves the best performance on K562 and HepG2 for the reported metrics. On SK-N-SH, Enformer remains stronger, suggesting that this cell type may require either richer priors in RCC or higher-quality/self-consistent rationales.
Importantly, R3LM provides mechanistic rationales alongside scores, enabling evidence-grounded inspection that is not available in black-box predictors.

We also report an \emph{oracle-gated} upper bound in which we keep only training examples where the generator's discretized level matches the ground truth. While this filtering is not used in our main method (because it conditions on labels), it quantifies the potential headroom from improving rationale quality: when rationales are consistently label-aligned, the same regressor architecture can reach substantially higher correlation.

\subsection{Study on Stage-1 Reasoning Generator}
\label{sec:cot_diagnostics}

Although R3LM ultimately performs \emph{continuous} regression on $y$, the stage-1 generator is trained to produce rationales and a discretized activity level $\ell$ (Section~\ref{sec:training}). We evaluate the generator on the large training split (excluding the rationale-supervised subset) to quantify structural reliability and level-prediction difficulty.

Table~\ref{tab:stage1_stats} shows that the generator achieves high syntactic validity (format pass rate $\approx 96$--$99\%$ across cell types), confirming that RCC and our constrained rationale format yield stable outputs. Exact level accuracy is around $0.42$, with mean absolute error below $1$ level on average. Notably, despite this imperfect level prediction, the downstream regressor remains strong under Generated-CoT evaluation (Table~\ref{tab:main_results}), indicating that R3LM can exploit partially correct mechanistic traces and RCC evidence for continuous reward prediction.

Figure~\ref{fig:analyze} further provides distributional diagnostics for the stage-1 generator and the resulting rationales.
Beyond exact-level accuracy, we report the \emph{within-one} criterion $\mathrm{Acc@1}=\Pr(|\Delta|\le 1)$, where $\Delta=\hat{\ell}-\ell$.
As shown in Fig.~\ref{fig:analyze}(a), prediction errors are strongly concentrated around zero, and $\mathrm{Acc@1}$ reaches 78.58\% (K562), 78.37\% (SKNSH), and 81.14\% (HepG2).
This indicates that most failures are \emph{near-miss} level shifts rather than large deviations, which helps explain why the downstream regressor remains competitive under the Generated-CoT setting (Table~\ref{tab:main_results})---it can still benefit from partially correct mechanistic traces and RCC-grounded evidence even when the discretized level is imperfect.

We also analyze the structural complexity and groundedness of CRE-ReasonBench rationales.
Fig.~\ref{fig:analyze}(b) summarizes the distribution of motif counts scanned into the RCC prompt, reflecting comparable prompt complexity across cell types.
Fig.~\ref{fig:analyze}(c) shows the distribution of rationale length (in tokens), indicating stable generation behavior under our constrained ``Step 1--7'' format.
Finally, Fig.~\ref{fig:analyze}(d) quantifies groundedness via \emph{motif coverage}, defined as $|M_{\mathrm{prompt}}\cap M_{\mathrm{reason}}|/|M_{\mathrm{prompt}}|$, i.e., the fraction of RCC prompt motifs explicitly referenced in the rationale.
The consistently high coverage suggests that the generator tends to cite motif evidence present in RCC rather than producing ungrounded narratives, supporting the interpretability claims of R3LM.

\begin{table}[t]
\centering
\small
\begin{tabular}{lccc}
\toprule
\textbf{Metric} & \textbf{K562} & \textbf{HepG2} & \textbf{SK-N-SH} \\
\midrule
Format-valid rate (\%) & 98.81 & 97.82 & 96.43 \\
Level accuracy (\%) & 41.84 & 43.23 & 42.12 \\
Level MAE ($\downarrow$) & 0.8256 & 0.7772 & 0.8196 \\
\bottomrule
\end{tabular}
\caption{\textbf{Stage-1 generator diagnostics.} Format validity and discretized level prediction on the large training split (per cell type).}
\label{tab:stage1_stats}
\end{table}

\subsection{Ablations and Design Choices}
\label{sec:ablations}

We conduct targeted ablations to isolate the impact of the RCC interface and implementation choices for continuous activity with LLM backbones.

\textbf{RCC vs.\ raw sequence for direct regression.}
We first fine-tune the LLM to directly regress activity scores from prompts \emph{without} explicit reasoning. Providing only the raw nucleotide string yields $r=0.3919$ on K562, whereas replacing the raw sequence-only prompt with RCC improves performance to $r=0.6886$. This supports our claim that ``compiling'' DNA into structured, grounded evidence is beneficial  before introducing reasoning traces.

\textbf{Anchor representation for score regression.}
We also compare different anchor strategies for extracting the representation used by the regression head. Using the last non-padding token representation performs best in our setting ($r=0.6886$), while introducing an explicit special token \texttt{<REG>} ($r=0.6537$) or using the EOS token ($r=0.5943$) is worse. We attribute this to the stability of the last-token representation under our formatting, which consistently aggregates the model's final hidden summary of the conditioning context.

\begin{table}[t]
\centering
\small
\begin{tabular}{lcc}
\toprule
\textbf{Ablation (K562)} & \textbf{Setting} & \textbf{Pearson $rho$} \\
\midrule
Direct regression & raw sequence prompt & 0.3919 \\
Direct regression & RCC prompt & 0.6886 \\
\midrule
Anchor choice & last non-pad token & 0.6886 \\
Anchor choice & \texttt{<REG>} special token & 0.6537 \\
Anchor choice & EOS token & 0.5943 \\
\bottomrule
\end{tabular}
\caption{\textbf{Ablations on K562 (Pearson $rho$).} RCC improves direct regression relative to raw sequence prompts. For score regression, the last non-padding token is a stronger anchor than inserting a dedicated special token or using EOS.}
\label{tab:ablations}
\end{table}

\subsection{Additional Evaluations and Sanity Checks}
\label{sec:additional_evals}

To further substantiate that R3LM performs \emph{regression based on reasoning} (rather than ignoring the rationale) and to strengthen robustness claims for reward modeling, we include the following additional diagnostics.

\textbf{Rationale shuffling at test time.}
In the Generated-CoT setting, we randomly permute rationales across test samples while keeping RCC prompts fixed. If the regressor truly relies on the rationale, performance should drop substantially. We observe a decrease from $r=\text{0.8246}$ to $r=\text{0.0198}$ on K562. This provides direct evidence that the rationale acts as an informative intermediate factor.
\section{Conclusion}
In this paper, we present R3LM to address the representation mismatch where raw DNA strings alone serve as an inefficient interface for mechanistic reasoning with LLMs. By compiling sequences into RCC—incorporating explicit motif evidence and experimental context—our approach enables LLMs to condition continuous regression on structured, auditable reasoning traces. Furthermore, we demonstrate that training on self-generated rationales bridges the inference gap, encouraging the model to rely on explicit, biologically grounded predictive features. Overall, R3LM outperforms strong sequence-based baselines across diverse cell types, achieving state-of-the-art accuracy while grounding predictions in human-inspectable biological mechanisms.

\section{Interdisciplinary Collaboration and Author Contributions}

This work represents an interdisciplinary collaboration between artificial intelligence and biology. Y.D., Z.Y., J.Z., Y.B., and B.S. brought expertise in machine learning, while C.C. contributed domain knowledge in biology and genomics.

B.S. conceptualized the study. C.C. provided biological guidance on regulatory mechanisms. Y.D. and Z.Y. designed the detailed method. Y.D. conducted experiments. Y.D. and  Z.Y. analyzed the results. Y.D., Z.Y., and B.S. wrote the draft. All authors participated in discussions throughout the project and reviewed the final manuscript.

\section{Limitations and Ethical Considerations}

\textbf{Limitations.} (1) \textit{Dependence on Priors:} Performance is bounded by the coverage of motif databases used in RCC construction; extending to de novo motif discovery is a future direction. (2) \textit{Scope:} We validate on MPRA datasets; generalization to endogenous chromatin or distal regulation requires further study. (3) \textit{Interpretability:} While reasoning traces improve transparency, they are model-generated proxies and should be treated as heuristic explanations rather than causal proofs.

\textbf{Ethical Considerations.} We use anonymized public data. Potential risks involve biases in training data or motif references, which could affect model fairness across different cell types. Outputs are intended for research screening, not clinical diagnosis.

\section{GenAI Disclosure}
This work makes explicit use of large language models (LLMs) as experimental components and data-generation tools.

First, in the preliminary probe study (Section 3.1), we queried GPT-5.2 via the OpenAI API to evaluate whether an instruction-tuned frontier LLM can infer regulatory properties directly from raw DNA sequences under controlled pairwise settings. These probes are used solely for analysis of representation limitations and are not part of the training data for any model evaluated in this paper.

Second, to construct CRE-ReasonBench (Section 3.3), we used the frontier LLM Claude-sonnet-4-5-20250929 to synthesize mechanistic rationales conditioned on structured Regulatory Context Card (RCC) prompts and observed discrete activity levels. These rationales are used as reasoning supervision and as oracle references for evaluation. The synthesis process follows a fixed, constrained instruction protocol enforcing structured, evidence-grounded reasoning. Outputs are filtered only for syntactic and structural validity (e.g., required fields and step structure), not for correctness with respect to labels beyond format compliance. A subset of the generated rationales was manually reviewed and corrected to form the supervised training set for the stage-1 reasoning generator.

Third, during model training and evaluation, LLMs are used as backbone architectures for fine-tuning and regression, but no external LLM calls are made at inference time beyond the trained models themselves.

Finally, LLMs were also used to assist with language polishing and clarity of presentation during manuscript preparation.

We emphasize that the mechanistic rationales in CRE-ReasonBench are synthesized explanations rather than ground-truth causal annotations. Conclusions regarding interpretability and robustness are therefore conditioned on the quality and assumptions of these generated rationales, and should be interpreted accordingly.

\begin{acks}

This work was supported in part by the National
Natural Science Foundation of China No. 62376277, Public Computing
Cloud, Renmin University of China, and fund for building world-class universities (disciplines) of Renmin University of China.
\end{acks}

\bibliographystyle{ACM-Reference-Format}
\bibliography{sample-base}

\appendix

\section{Training Details}
\label{app:training_details}
\textbf{Data split.}
All training/validation/test samples follow the official regLM split.

\textbf{Stage-I supervised fine-tuning.}
For each cell type, we curate 1{,}000 instruction-following training instances generated by Claude-sonnet-4-5-20250929 and subsequently human-reviewed and corrected.
We conduct full-parameter fine-tuning using the LLaMA-Factory framework\cite{zheng2024llamafactory}, initialized from \texttt{Qwen3-4B-Instruct-2507}\cite{qwen3technicalreport}.
Unless otherwise specified, we use the following hyperparameters:
\begin{itemize}
    \item \texttt{num\_train\_epochs}: 100
    \item \texttt{per\_device\_train\_batch\_size}: 1
    \item \texttt{gradient\_accumulation\_steps}: 16
    \item \texttt{learning\_rate}: $1.0\times 10^{-5}$
    \item \texttt{lr\_scheduler\_type}: cosine
    \item \texttt{warmup\_ratio}: 0.1
    \item \texttt{optim}: \texttt{adamw\_torch}
    \item \texttt{weight\_decay}: 0.01
    \item \texttt{bf16}: true
    \item \texttt{deepspeed}: \texttt{ZeRO2}
\end{itemize}
Stage~I training runs on 8$\times$ NVIDIA A100-SXM4-80GB GPUs and takes approximately 2 hours.

\textbf{Offline generation for rationale construction.}
All offline generations are produced with vLLM using default settings, with decoding parameters:
\texttt{temperature}=0.95,
\texttt{top\_p}=0.7,
\texttt{top\_k}=50,
\texttt{max\_new\_tokens}=1024,
and \texttt{repetition\_penalty}=1.0.

\textbf{Stage-II regression training.}
Stage~II is implemented in a custom PyTorch Lightning pipeline and initialized from \texttt{Qwen3-4B
-Instruct-2507}.
We optimize the regression objective for up to 10 epochs with:
\begin{itemize}
    \item \texttt{lr}: $1.0\times 10^{-5}$
    \item \texttt{weight\_decay}: 0.01
    \item \texttt{max\_epochs}: 10
    \item \texttt{batch\_size}: 2
    \item \texttt{gradient\_accumulation\_steps}: 4
    \item \texttt{warmup\_ratio}: 0.1
    \item \texttt{max\_grad\_norm}: 200.0
    \item \texttt{bf16}: true
\end{itemize}
We enable LoRA on the backbone with \texttt{lora\_r}=16, \texttt{lora\_alpha}=32, and \texttt{lora\_dropout}=0.05.
Stage~II training runs on 8$\times$ NVIDIA A100-SXM4-80GB GPUs and takes approximately 16 hours.

\section{Why RCC helps}
\label{app:why_rcc}
RCC is designed to support scientific learning in three complementary ways.
\textit{(i) Comparability:} each sample is represented by an isomorphic field structure, enabling cross-sequence alignment and systematic perturbations under a consistent schema.
\textit{(ii) Groundedness:} each field is produced by traceable algorithms (motif scans, statistics, explicit heuristics), making the evidence auditable and limiting purely linguistic confabulations.
\textit{(iii) Reasoning-friendliness:} by surfacing motifs, grammar hypotheses, and context as explicit intermediate factors, RCC encourages models to construct stepwise mechanistic explanations (e.g., ``Step 1--Step 7'') rather than implicitly inferring regulatory structure from next-token statistics over raw DNA strings.
In the next section, we leverage RCC to synthesize grounded mechanistic rationales at scale, yielding a supervision signal suitable for training a reasoning-conditioned continuous reward model.

\section{Case Study: A long-distance \textit{NBEAL2} enhancer in K562 supports grounded CoT reasoning}
\label{app:case-nbeal2}

\textbf{Literature background.}
Gray \emph{et al.} identified a regulatory region located $\sim$31 kb upstream of \textit{NBEAL2} (within an intron of \textit{CCDC12}) as an enhancer locus in K562 \cite{gray_gata1_nbeal2_enhancer}. The region is marked by enhancer-associated chromatin signals (e.g., H3K4me1 and H3K27ac) and DNase hypersensitivity, shows GATA1/GATA2 ChIP-seq occupancy, and is predicted to physically interact with the \textit{NBEAL2} promoter (ChIA-PET). Reporter assays in K562 confirmed enhancer activity for two overlapping fragments (BS-1-2-3 and BS-4-5). Crucially, site-directed mutagenesis demonstrated that disrupting \textbf{GATA binding sites BS-2 or BS-3} strongly reduces (and can block) enhancer activity, whereas mutating BS-1/BS-4/BS-5 has little effect. DNA-binding assays further showed that GATA1 and GATA2 bind the enhancer fragments, and GATA1 knockdown in K562 decreases \textit{NBEAL2} expression.

\textbf{Model input and prediction.}
Given the wild-type enhancer sequence, our RCC compilation (motif evidence, structural cues, and context) yielded a high GC content (0.586) and prominent GATA-family motif hits near the experimentally validated BS-2/BS-3 region. Conditioned on this RCC, R3LM generated the following structured reasoning trace and predicted \textbf{Activity level 2 (Moderate Activity)}.

\textbf{Model reasoning trace (wild-type).}
\begin{quote}\small
\textbf{Step 1:} GC content analysis --- 0.586 indicates high GC content ($>0.55$), correlating with open chromatin in K562 and strong enhancer potential.\\
\textbf{Step 2:} GATA1/2 dependency validation --- the sequence is a GATA1/2-dependent enhancer with critical BS-2/BS-3 sites, implying spatial requirements for GATA occupancy.\\
\textbf{Step 3:} Functional site analysis --- BS-2 and BS-3 match the canonical GATA motif (WGATAR) and act as primary recruitment platforms; other sites are auxiliary.\\
\textbf{Step 4:} Partner factor recruitment --- proposes possible erythroid co-factor assembly near BS-2/BS-3 (e.g., TAL1/KLF1/FOG1), consistent with canonical GATA enhancer logic.\\
\textbf{Step 5:} Chromatin context integration --- predicts enhancer-like marks (e.g., H3K4me1/H3K27ac) and accessibility enabling co-activator recruitment.\\
\textbf{Step 6:} Activity reconciliation --- predicts moderate activity, hypothesizing constraints such as incomplete co-factor assembly or suboptimal motif grammar.\\
\textbf{Step 7:} Final determination --- high accessibility + BS-2/BS-3 dependence supports strong regulatory potential but consistent with a moderate output state (Level 2).
\end{quote}

\textbf{Why this case supports ``good'' CoT (qualitative alignment).}
This case study provides a direct, literature-grounded check of the generated reasoning:
(i) \textbf{Key-causal site identification:} the trace explicitly highlights BS-2/BS-3 as the functional core (Steps 2--3), matching the mutagenesis result that BS-2/BS-3 disruptions strongly reduce enhancer activity.
(ii) \textbf{Grounded chromatin-context claims:} although the model used GC content as a heuristic for accessibility (Step 1), the paper independently reports DNase hypersensitivity and enhancer-associated histone marks at this locus (Step 5).
(iii) \textbf{Mechanistic plausibility beyond correlation:} the reasoning invokes a testable, site-centric mechanism (GATA occupancy at BS-2/BS-3 driving enhancer output) that the paper supports via luciferase assays, DNA-binding assays, and GATA1 knockdown affecting \textit{NBEAL2} expression.
Finally, the model's ``Moderate Activity'' label is compatible with the fact that luciferase assays and MPRA-derived activity levels use different experimental scales; our trace (Step 6) explicitly acknowledges that regulatory potential and measured output can differ across cellular states and assay contexts.

\textbf{Takeaway.}
Overall, the CoT is not a free-form narrative: it selects the same decisive binding sites validated by targeted mutagenesis, anchors accessibility claims to enhancer-like chromatin context, and yields a coherent, falsifiable hypothesis for enhancer function in K562.
\end{document}